\documentclass[sigconf]{aamas}
\pdfoutput=1

\usepackage{balance}
\usepackage[utf8]{inputenc}
\usepackage{caption}
\usepackage{xcolor}
\usepackage{xspace}
\usepackage{amsthm}
\usepackage{amsmath}
\usepackage{amsfonts}

\usepackage{tikz}
\usetikzlibrary{arrows,automata,shapes,backgrounds,fit,positioning}

\usepackage{hyperref}
\hypersetup{
    colorlinks,
    linktocpage,
    citecolor=black,
    filecolor=black,
    linkcolor=blue,
    urlcolor=black,
}

\setcopyright{ifaamas}
\acmConference[AAMAS '21]{Proc.\@ of the 20th International Conference on Autonomous Agents and Multiagent Systems (AAMAS 2021)}{May 3--7, 2021}{Online}{U.~Endriss, A.~Now\'{e}, F.~Dignum, A.~Lomuscio (eds.)}
\copyrightyear{2021}
\acmYear{2021}
\acmDOI{}
\acmPrice{}
\acmISBN{}

\newcommand{\lan}[1]{\ensuremath{\mathbf{#1}}\xspace}
\newcommand{\stratstyle}[1]{\ensuremath{\mathrm{#1}}}
\newcommand{\ATLext}[1]{\lan{ATL_\stratstyle{#1}}}
\newcommand{\ATL}{\ATLext{}}

\newcommand{\coop}[2][]{\langle\!\langle{#2}\rangle\!\rangle_{_{\!\mathit{#1}}}}
\newcommand{\complexityclass}[1]{\ensuremath{\mathbf{{#1}}}\xspace}
\newcommand{\Ptime}{\complexityclass{P}}

\newcommand{\Pspace}{\complexityclass{PSPACE}}

\newcommand{\Exptime}{\complexityclass{EXPTIME}}
\newcommand{\EXPTIME}{\Exptime}
\newcommand{\Deltacomplx}[1]{\complexityclass{\Delta_{{#1}}^{\Ptime}}}
\newcommand{\Deltwo}{\Deltacomplx{2}}
\newcommand{\Sometm}{\mathrm{F}}
\newcommand{\Always}{\mathrm{G}}
\newcommand{\prop}[1]{\textsf{#1}}
\newcommand{\tool}{\textbf{STV+Reductions}\xspace}
\newcommand{\para}[1]{\smallskip\noindent\textit{#1}}

\title[\tool]{\tool: Towards Practical Verification \\ of Strategic Ability Using Model Reductions}
\subtitle{Demonstration Track}

\author{Damian Kurpiewski}
\affiliation{
  \institution{Institute of Computer Science, Polish Academy of Sciences}}
\email{d.kurpiewski@ipipan.waw.pl}

\author{Witold Pazderski}
\affiliation{
  \institution{Institute of Computer Science, Polish Academy of Sciences}}
\email{w.pazderski@ipipan.waw.pl}

\author{Wojciech Jamroga}
\affiliation{
  \institution{Institute of Computer Science, Polish Academy of Sciences and University of Luxembourg}}
\email{w.jamroga@ipipan.waw.pl}

\author{Yan Kim}
\affiliation{
  \institution{University of Luxembourg}}
\email{yan.kim@uni.lu}

\begin{abstract}
  We present a substantially expanded version of our tool \textbf{STV} for strategy synthesis and verification of strategic abilities. The new version adds user-definable models and support for model reduction through partial order reduction and checking for bisimulation.
\end{abstract}


\begin{document}

\pagestyle{fancy}
\fancyhead{}

\maketitle

\section{Introduction}\label{sec:intro}

Formal analysis of multi-agent systems is becoming increasingly important as the procedures, protocols, and technology that surround us get more and more complex.
\emph{Alter\-nating-time temporal logic} \ATL~\cite{Alur97ATL,Alur02ATL,Schobbens04ATL} is probably the most popular logic to describe interaction in MAS.
Formulas of \ATL allow to express statements about what agents (or groups of agents) can achieve.
For example, $\coop{taxi}\Always\neg\prop{fatality}$ says that the autonomous cab can drive in such a way that nobody is ever killed, and $\coop{taxi,passg}\Sometm\,\prop{destination}$ expresses that the cab and the passenger have a joint strategy to arrive at the destination, no matter what any other agents do.

Algorithms and tools for verification of such properties have been in development for over 20 years~\cite{Alur98mocha-cav,Alur01jmocha,Kacprzak04umc-atl,Lomuscio06mcmas,Chen13prismgames,Busard14improving,Pilecki14synthesis,Huang14symbolic-epist,%
Cermak14mcheckSL,Cermak15mcmas-sl-one-goal,Lomuscio17mcmas,Belardinelli17broadcasting,Belardinelli17publicActions,Jamroga19fixpApprox-aij,Kurpiewski19stv-demo}.
Unfortunately, model checking of agents with imperfect information is  \Deltwo- to \Pspace-complete for memoryless strategies~\cite{Schobbens04ATL,Jamroga06atlir-eumas,Bulling10verification} and \EXPTIME-complete to undecidable for agents with perfect recall~\cite{Dima11undecidable,Guelev11atl-distrknowldge}; also, the problem does not admit simple incremental solutions~\cite{Bulling11mu-ijcai,Dima14mucalc,Dima15fallmu}.
This has been confirmed in experiments~\cite{Pilecki14synthesis,Busard14improving,Huang14symbolic-epist,Busard15reasoning,Jamroga19fixpApprox-aij} and case studies~\cite{Jamroga18Selene,Jamroga20Pret-Uppaal,Jamroga20natvoting}.

Much of the complexity is due to the size of the model, and in particular to {state space explosion}~\cite{Baier08mcheck}.
To address the problem, we have extended our experimental tool \textbf{STV (STrategic Verifier)}~\cite{Kurpiewski19stv-demo} with support for \emph{model reductions}.
Two methods are used: (i) checking for equivalence of models according to a handcrafted relation of \emph{$A$-bisimulation}~\cite{Jamroga21Bisimulations}, and (ii) fully automated \emph{partial order reduction (POR)}~\cite{Jamroga18por,Jamroga20POR-JAIR}.
We also add a simple model specification language that allows the user to define their own inputs for verification, which was not available in the previous version~\cite{Kurpiewski19stv-demo}.

The purpose of the extension is twofold. First, it should facilitate practical verification of MAS, as the theoretical and experimental results for POR and bisimulation-based reduction suggest~\cite{Jamroga20POR-JAIR,Jamroga21Bisimulations}.
No less importantly, it serves a pedagogical objective, as we put emphasis on visualisation of the reductions, so that the tool can be also used in the classroom to show how the reduction works.
Finally, checking strategic bisimulation by hand is difficult and prone to errors; here, the user can both see the idea of the bisimulation, and automatically check if it is indeed correct.

\section{Application Domain}\label{sec:domain}

\tool is aimed at verification of agents' abilities -- in particular, synthesis of memoryless imperfect information strategies that guarantee a given temporal goal.
This includes both model checking of \emph{functionality requirements} (understood as the ability of legitimate users to achieve their goals), and \emph{security properties} defined by the inability of an intruder to compromise the system.

A good example of a specific domain is formal verification of voting procedures and elections, with a number of classical requirements, such as
\emph{election integrity}, \emph{ballot secrecy}, \emph{receipt-freeness}, and \emph{voter-verifiability}~\cite{Ryan10atemyvote,Tabatabaei16expressing}.
Some recent case studies~\cite{Jamroga18Selene,Jamroga20Pret-Uppaal,Jamroga20natvoting} have shown that practical verification of such scenarios is still outside of reach.
Some tools do not support intuitive specification and validation of models; some others have limited property specification languages.
In all cases, the state-space explosion is a major obstacle that prevents verification of anything but toy models.

\section{Scenarios}\label{sec:scen}

The new version of \textbf{STV} provides a flexible specification language for asynchronous models.
The following examples are included:
Train-Gate-Controller (TGC)~\cite{Alur98mocha-cav,Hoek02ATEL,Jamroga20POR-JAIR}, Two-Stage Voting~\cite{Jamroga21Bisimulations}, and Asynchronous Simple Voting~\cite{Jamroga20POR-JAIR}.
Some built-in synchronous models are also included, such as TianJi~\cite{Lomuscio17mcmas}, Castles~\cite{Pilecki17smc}, Bridge Endplay~\cite{Jamroga17fixpApprox}, and Drones~\cite{Jamroga20mvATL}.

\begin{figure}
\begin{tikzpicture}[->,>=stealth',shorten >=1pt,auto,node distance=1.4cm,semithick,inner sep=2pt,bend angle=45]

\tikzstyle{every node}=[font=\tiny]

\node[state] (f0) [fill=blue!20] {$G, W, W$};
\node[state] (f1) [below left of=f0, fill=blue!20] {$R, T, W$};
\node[state] (f2) [below right of=f0, fill=blue!20] {$R, W, T$};
\node[state] (f3) [below of=f1, fill=blue!20] {$G, A, W$};
\node[state] (f4) [below of=f2, fill=blue!20] {$G, W, A$};
\node[state] (f5) [below of=f3] {$R, A, T$};
\node[state] (f6) [below of=f4] {$R, T, A$};
\node[state] (f7) [below right of=f5] {$G, A, A$};

\tikzstyle{every node}=[font=\normalsize]

\node[state, scale=0.8] (W1) [left of=f1,node distance=6cm] {$W$};
\node[state, scale=0.8] (T1) [below of=W1] {$T$};
\node[state, scale=0.8] (A1) [below of=T1] {$A$};

\node[state, scale=0.8] (G)  [right of=W1] {$G$};
\node[state, scale=0.8] (R)  [below of=G] {$R$};

\node[state, scale=0.8] (W2) [right of=G] {$W$};
\node[state, scale=0.8] (T2) [below of=W2] {$T$};
\node[state, scale=0.8] (A2) [below of=T2] {$A$};

\node[] (Train1) [below of=A1,node distance=1cm,font=\scriptsize] {$Train1$};
\node[] (Controller) [right of=Train1,node distance=1.15cm,font=\scriptsize] {$Controller$};
\node[] (Train2) [below of=A2,node distance=1cm,font=\scriptsize] {$Train2$};

\tikzstyle{every node}=[font=\footnotesize]

\path (W1) edge              node [left] {$a_1$} (T1)
			(T1) edge              node [left] {$a_2$} (A1)
			(A1) edge [bend left,near end]  node {$a_3$} (W1)
			(W2) edge              node {$b_1$} (T2)
			(T2) edge              node {$b_2$} (A2)
			(A2) edge [bend right,near end]  node [right] {$b_3$} (W2)
			(G)  edge [bend right=20]  node [left] {$a_1$} (R)
			     edge [bend left=20] node {$b_1$} (R)
			(R)  edge [bend right=90]  node [right] {$b_2$} (G)
			     edge [bend left=90] node {$a_2$} (G);
					
\tikzstyle{every node}=[font=\tiny]

\path (f0) edge [blue] node [above] {$a_1$} (f1)
      (f0) edge [blue] node [above] {$b_1$} (f2)
			(f1) edge [blue] node [left] {$a_2$} (f3)
			(f2) edge [blue] node [right] {$b_2$} (f4)
			(f3) edge              node [left] {$b_1$} (f5)
			(f4) edge              node [right] {$a_1$} (f6)
			(f5) edge              node [left] {$b_2$} (f7)
			(f6) edge              node [right] {$a_1$} (f7)
			(f7) edge [bend left=90]  node [left] {$b_3$} (f3)
			(f7) edge [bend right=90]  node [right] {$a_3$} (f4)
			(f3) edge [bend left=90, blue]  node [left] {$a_3$} (f0)
			(f4) edge [bend right=90, blue]  node [right] {$b_3$} (f0)
			(f5) edge [bend right]  node [left] {$a_3$} (f1)
			(f6) edge [bend left]  node [right] {$b_3$} (f2);
                        	
\end{tikzpicture}
\caption{Trains, Gate, and Controller benchmark (TGC): asynchronous MAS (left); full and reduced model (right).}
\label{fig:tgc}
\end{figure}
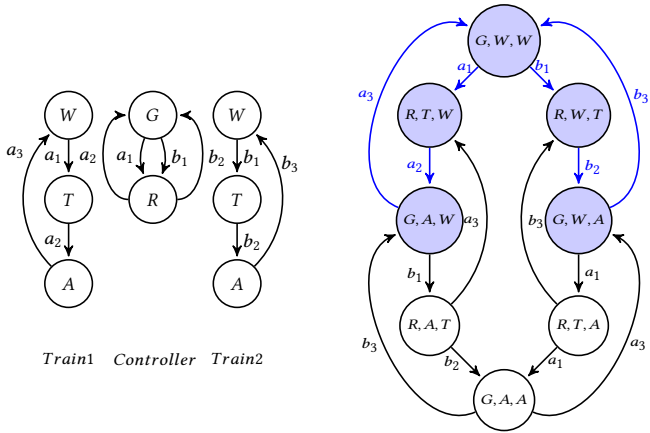

\section{Formal Background}\label{sec:prelim}

\para{Models.}
The main part of the input is given by an \emph{asynchronous multi-agent system (AMAS)}~\cite{lomuscio10partialOrder,Jamroga20POR-JAIR,Jamroga20paradoxes-tr}, i.e., a network of local automata (one automaton per agent).
From the AMAS, the global model is generated, where nodes are tuples of local states.
The knowledge/uncertainty of an agent is defined by the agent's local state.
An example AMAS is shown in Figure~\ref{fig:tgc}(left).
The global model generated from the AMAS is shown in Figure~\ref{fig:tgc}(right).

\nocite{Fagin95knowledge}

\para{Strategies.}
A strategy is a conditional plan that specifies what the agent(s) are going to do in every possible situation.
Here, we consider the case of \emph{imperfect information memoryless strategies}, represented by functions from the agent's local states (formally, abstraction
classes of its indistinguishability relations) to its available actions.
The \emph{outcome} of a strategy from state $q$ consists of all the infinite paths starting from $q$ and consistent with the strategy.

\para{Formulas.}
Given a model $M$ and a state $q$ in the model, the formula $\coop{A}\varphi$ holds in $(M,q)$ iff there exists a strategy for $A$ that makes $\varphi$ true on all the outcome paths starting from any state indistinguishable from $q$.
For more details, we refer the reader to~\cite{Alur02ATL,Schobbens04ATL}.

\para{Model reduction and bisimulation.}
State space explosion is a major factor that prevents practical model checking~\cite{Baier08mcheck}.
A possible way out is \emph{model reduction}, i.e., using a smaller equivalent model for verification instead of the original one.
A suitable notion of \emph{$A$-bisimulation} has been proposed in~\cite{Jamroga21Bisimulations}.
Unfortunately, synthesizing a reduced $A$-bisimilar model is at least as hard as the verification itself~\cite{Jamroga21Bisimulations}. However, checking if a handcrafted relation is an $A$-bisimulation can be done in polynomial time, which offers valuable help especially for larger models.

\para{Partial-order reduction.}
A fully automated model reduction is possible if the state space explosion is due to asynchronous interleaving of agents' actions.
The method is called \emph{partial order reduction}, and has been been recently extended to verification of strategic abilities under imperfect information~\cite{Jamroga20POR-JAIR}.
The reduced model for the TGC scenario is highlighted in blue color in Figure~\ref{fig:tgc}(right).

\section{Technology}\label{sec:tech}

\tool does \emph{explicit-state model checking}.
That is, the global states and transitions of the model are represented explicitly in the memory of the verification process.
The tool includes the following new functionalities.

\para{User-defined input.}
The user can load and parse the input specification from a text file that defines: the local automata in the AMAS, the formula to be verified, the propositional variables, persistent propositions, agent names relevant for POR, and/or the mapping for bisimulation checking.
Based on that, the global model is generated and displayed in the GUI and can be verified by means of \emph{fixpoint approximation}~\cite{Jamroga19fixpApprox-aij} or \emph{dominance-based strategy search}~\cite{Kurpiewski19domination}.
When using partial-order reduction, the reduced model is also displayed, and highlighted in the full model.

\para{Partial-order reduction.}
The fully automated reduction method is based on POR~\cite{Peled93representatives} and implemented according to the algorithms proposed in~\cite{LomuscioPQ10,Jamroga20POR-JAIR}.
The reduced model is generated based on the AMAS specification, together with two additional parameters: the coalition and the set of proposition variables.

\para{Bisimulation checking.}
The tool allows to check if two models are $A$-bisimilar for a given coalition $A$~\cite{Jamroga21Bisimulations}.
Apart from the specification of the two models, the bisimulation relation between the corresponding states must also be provided, along with the selected coalition.

\section{Usage}\label{sec:eval}

The current version of \tool is available for download \href{https://github.com/blackbat13/stv/releases/tag/v0.2-alpha}{\color{blue}here}, and allows to:
\begin{itemize}
\item Select and display a model specification from a text file,
\item Generate and display the explicit state-transition graph,
\item Generate and display the reduced model using POR,
\item Select specifications of two models and a relation from text files, and check if the models are $A$-bisimilar wrt the relation,
\item Verify the selected full or reduced model by means of fixpoint approximation or dominance-based verification (DominoDFS),
\item Alternatively, run the verification for a predefined parameterized model and formula,
\item Display the verification result, including the relevant truth values and the winning strategy.
\end{itemize}

\section{Conclusions}\label{sec:conclusions}

Model checking strategic abilities under imperfect information is notoriously hard.
\tool addresses the state explosion problem by an implementation of partial-order reduction and bisimulation checking.
This should not only facilitate verification, but also make the techniques easier to use and understand.

\para{Acknowledgements.}
The authors acknowledge the support of the Luxembourg National Research Fund (FNR) and the National Centre for Research and Development (NCBiR), Poland, under the \\ CORE/PolLux project STV (POLLUX-VII/1/2019).

\newpage

\balance

\end{document}